\documentclass[aps,preprint,prd,showpacs,nofootinbib]{revtex4}

\usepackage{amsmath}
\usepackage{graphicx}
\usepackage{subfigure}
\usepackage{multirow}
\usepackage{bm}
\usepackage{amssymb}
\usepackage{mathtools}
\usepackage{enumerate}
\usepackage{color}
\usepackage[colorlinks,linkcolor=magenta,anchorcolor=cyan,citecolor=blue]{hyperref}

\def\be{\begin{equation}}
\def\ee{\end{equation}}

\begin{document}
\title{Implication of the Hubble tension for the primordial Universe in light of
recent cosmological data}

\author{Gen Ye$^{1}$\footnote{yegen14@mails.ucas.ac.cn}}
\author{Bin Hu$^{2}$\footnote{bhu@bnu.edu.cn}}
\author{Yun-Song Piao$^{1,3,4,5}$\footnote{yspiao@ucas.ac.cn}}

\affiliation{$^1$ School of Physics, University of Chinese Academy of
    Sciences, Beijing 100049, China}

\affiliation{$^2$ Department of Astronomy, Beijing Normal University, Beijing 100875, China}

\affiliation{$^3$ School of Fundamental Physics and Mathematical
    Sciences, Hangzhou Institute for Advanced Study, UCAS, Hangzhou
    310024, China}

\affiliation{$^4$ International Center for Theoretical Physics
    Asia-Pacific, Beijing/Hangzhou, China}

\affiliation{$^5$ Institute of Theoretical Physics, Chinese
    Academy of Sciences, P.O. Box 2735, Beijing 100190, China}

\begin{abstract}

In prerecombination resolutions of the Hubble tension, such as
early dark energy, new physics before recombination shifts the
values of relevant cosmological parameters so that the models can
fit with cosmic microwave background and baryon acoustic oscillations observations as well as $\Lambda$CDM does. In this
paper, we clarify how the parameter shifts are related with
$\delta H_0$, particularly we find the shift of primordial scalar
spectral index scales as ${\delta n_s}\simeq 0.4{\delta H_0\over
H_0}$ by performing the Monte Carlo Markov chain analysis with
the Planck2018+BAO+Pantheon+R19+Keck Array/BICEP dataset. A novel
point of our result is that if the current $H_0$ measured locally
is correct, complete resolution of the Hubble tension seems to be
pointing to a scale invariant Harrison-Zeldovich spectrum, i.e.
$n_s= 1$ for $H_0\sim 73$km/s/Mpc.

\end{abstract}
\maketitle
\section{Introduction}\label{sec:intro}

The Hubble constant $H_0$ quantifies the current expansion rate of
our Universe. It can be predicted based on observations of
anisotropies in the cosmic microwave background (CMB) and other
early universe physics such as baryon acoustic oscillations (BAO).
Assuming the standard cosmological model ($\Lambda$CDM), the
Planck collaboration has reported $H_0=67.4\pm0.5$km/s/Mpc
\cite{Aghanim:2018eyx}. Recently, $H_0$ has been also measured by
lots of local observations (up to percent level accuracy).
However, almost all yield $H_0\sim 73$km/s/Mpc, which is in
stark ($>4\sigma$) tension with that reported by Planck
collaboration based on the $\Lambda$CDM model \cite{Verde:2019ivm,
Riess:2020sih}, usually dubbed the Hubble tension. The origin of
this tension is still under investigation, but it is unlikely to
be explained by unknown systematic errors \cite{Bernal:2016gxb,
Feeney:2017sgx, Aylor:2018drw}.

Recently, it has been widely thought that the Hubble tension is
suggesting new physics beyond $\Lambda$CDM \cite{Verde:2019ivm,
Riess:2020sih,Knox:2019rjx,DiValentino:2020zio,Lyu:2020lwm,Haridasu:2020pms},
see Ref.\cite{DiValentino:2021izs} for a thorough review on
various ideas, see also, e.g, Ref.\cite{AresteSalo:2021wgb,AresteSalo:2021lmp,Dainotti:2021pqg} for some more recent discussions. Reducing the sound horizon $r^*_s=\int_{z_*}^\infty
c_s/H(z) dz$ ($z_*$ is the redshift at recombination), as in early
dark energy \cite{Poulin:2018cxd, Agrawal:2019lmo, Lin:2019qug,
Ye:2020btb, Alexander:2019rsc, Smith:2019ihp, Niedermann:2019olb,
Sakstein:2019fmf, Chudaykin:2020acu, Ye:2020oix, Braglia:2020iik,
Lin:2020jcb, Niedermann:2020dwg,
Chudaykin:2020igl,Fujita:2020ecn,Seto:2021xua,Tian:2021omz,Sabla:2021nfy,Nojiri:2021dze}
(non-negligible only for a short epoch decades before
recombination) or early modified gravity models \cite{
Zumalacarregui:2020cjh,
Ballesteros:2020sik,Braglia:2020bym,Braglia:2020auw,Odintsov:2020qzd},
is a promising road towards the complete resolution of the Hubble
tension. Since probes of the early universe, such as CMB and BAO,
set the angular scales $\theta^*_s\equiv r_s^*/D_A^*$ ($D_A^*\sim
1/H_0$ is the angular diameter to last scattering surface), a
smaller $r_s^*$ naturally brings a larger $H_0$. It should be
mentioned that the beyond-$\Lambda$CDM modifications after
recombination are difficult to reconcile with low redshift data
(light curves and BAO) \cite{Aylor:2018drw, Feeney:2017sgx,Lemos:2018smw, Efstathiou:2021ocp}; see
also e.g.
\cite{Vagnozzi:2019ezj,DiValentino:2019jae,Yang:2020zuk,Ye:2020btb,
yang20212021h0,Yang:2021eud}.

In corresponding early dark energy (EDE) models, $H_0\gtrsim
70$km/s/Mpc, moreover, the existence of anti de-Sitter (AdS) vacua
around recombination can further lift $H_0$ to $\sim
73$km/s/Mpc \cite{Ye:2020btb,Ye:2020oix}. Although many of the early
resolutions of the Hubble tension have been found to fit with CMB,
BAO and light curve observations as well as $\Lambda$CDM does, the cost
of compensating for the impact of new physics before recombination
is that the values of relevant parameters $\omega_{cdm}$,
$\omega_{b}$ and $n_s$ must be shifted \cite{Poulin:2018cxd,
Agrawal:2019lmo, Lin:2019qug, Ye:2020oix}. The parameter shifts
not only make the corresponding early resolution models tested by
upcoming CMB experiments, but also have potential implications to
the inflation and primordial Universe, see also
\cite{Benetti:2013wla,Gerbino:2016sgw,Zhang:2017epd,Benetti:2017gvm,Benetti:2017juy}
for relevant studies. Thus it is significant to have a full
insight into the shift patterns of parameters, specially the shift
of the spectral index $n_s$ (the primordial perturbation spectrum
$P_s\sim (k/k_{pivot})^{n_s}$).

\begin{figure}
\centering
\includegraphics[width=0.8\linewidth]{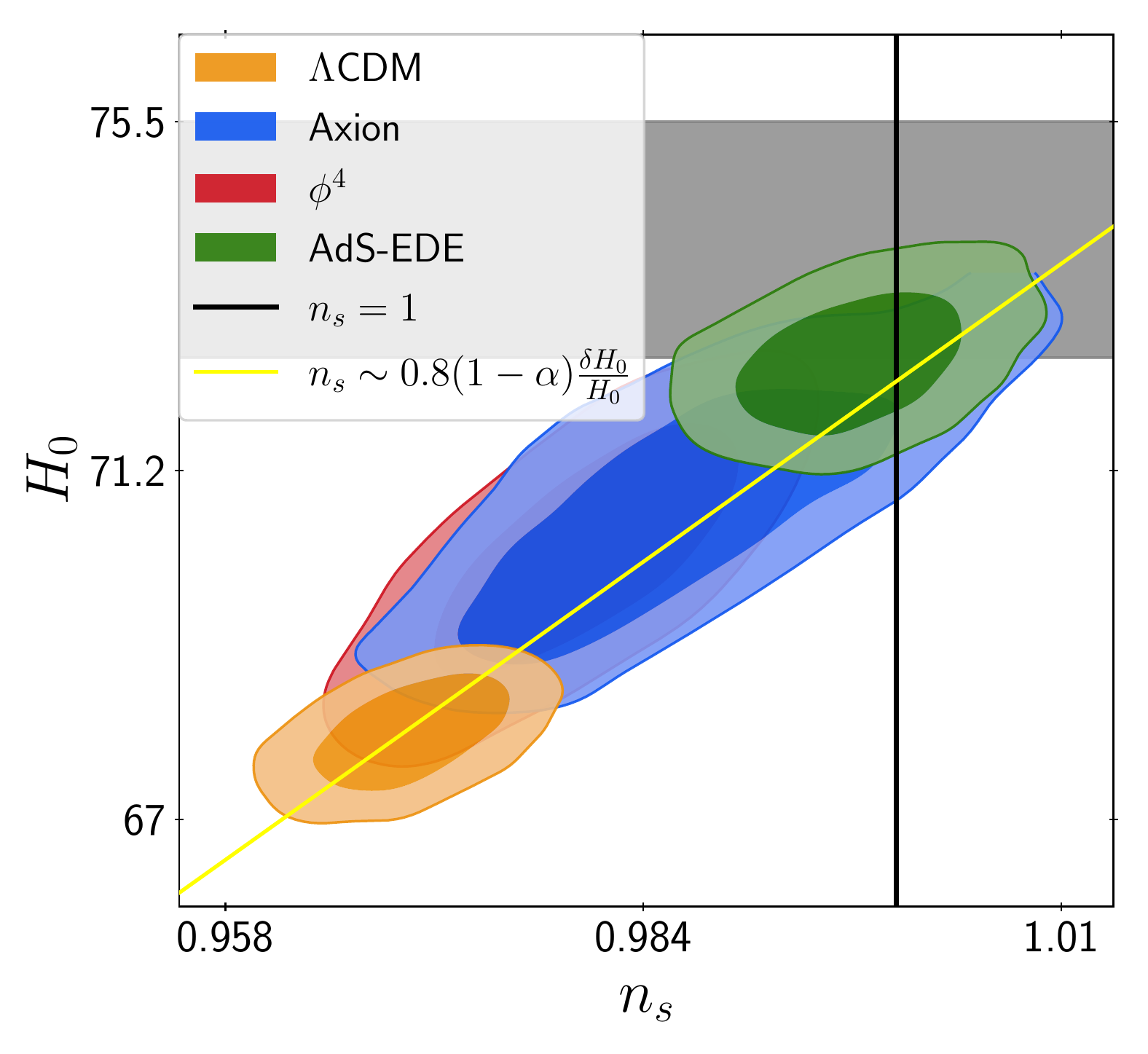}
\caption{The Hubble constant vs. $n_s$ plot. The EDE models:
the $n=3$ Axion model \cite{Poulin:2018cxd}, the $n=2$ Rock `n'
model \cite{Agrawal:2019lmo} (called $\phi^4$ for simplicity) and
the AdS-EDE model \cite{Ye:2020oix}. Dataset: Planck2018+Keck
Array/BICEP2015+BAO+Pantheon+R19, also for
Figs.\ref{r-ns},\ref{MID} and Tables \ref{ede tab}, \ref{chi2}. We
see that the $\Lambda$CDM and EDE models are consistent with the
$n_s$-$H_0$ scaling relation \eqref{ns-H0}. }\label{H0-ns}
\end{figure}

To identify the common pattern of parameter shifts in different
models, we first have to marginalize over the model-specific
information. We focus on the prerecombination resolutions
(referred to as early resolutions) of the Hubble tension satisfying the following:
\begin{itemize}
\item Reduce $r_s^*$ to lift $H_0$ ($r_s^*H_0\sim const.$).
    \item The evolution after recombination is described by
    $\Lambda$CDM.
\item The recombination process is not modified \footnote{There
are also proposals modifying the recombination process, e.g.
\cite{Chiang:2018xpn, Hart:2019dxi, Sekiguchi:2020teg}, see also
\cite{Jedamzik:2020krr} for the primordial magnetic fields.}.
\end{itemize}
We show how the parameter shifts in early resolution models
are related scalingly with $H_0$. Specially, for $n_s$, we get
\begin{equation}
{\delta n_s}\simeq 0.4{\delta H_0\over H_0}, \label{ns-H0}
\end{equation}
see also Fig.\ref{H0-ns} for the Monte Carlo Markov chain (MCMC)
analysis with joint Planck2018+BAO+Pantheon+R19 dataset, as well
as recent Keck Array/BICEP data \cite{Array:2015xqh}.
(\ref{ns-H0}) explains how the early resolutions of the Hubble
tension bring about a larger $n_s$ than $\Lambda$CDM, as observed
in e.g. Refs. \cite{Poulin:2018cxd, Agrawal:2019lmo, Lin:2019qug,
Ye:2020oix}, in which $n_s\gtrsim 0.98$ for $H_0\gtrsim
71$km/s/Mpc.

\begin{figure}
\centering
\includegraphics[width=0.8\linewidth]{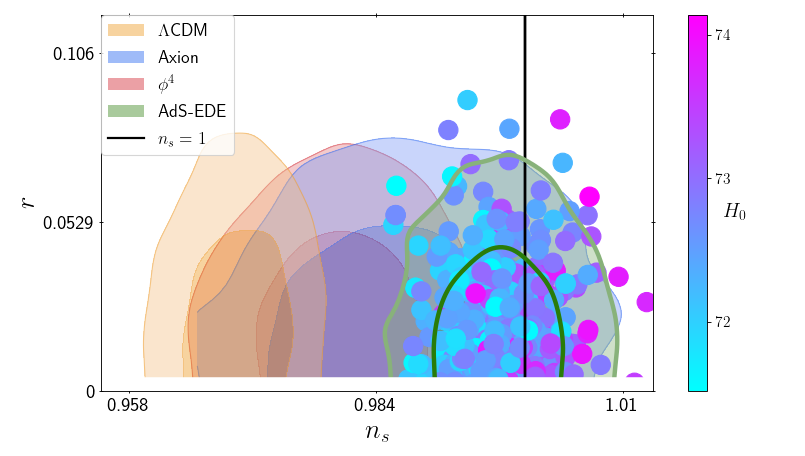}
\caption{The tensor-to-scalar ratio $r$ vs. $n_s$ plot. The pivot
scale $k_{pivot}=0.05\text{Mpc}^{-1}$. Scattered points correspond
to the AdS-EDE model with a color coding for $H_0$. As indicated
by the contours and scattered points, $n_s\simeq1$ for
$H_0\sim73$km/s/Mpc.} \label{r-ns}
\end{figure}

The exact scale invariant primordial spectrum ($n_s=1$), i.e. the
Harrison-Zeldovich spectrum proposed first in
\cite{Harrison:1969fb, Zeldovich:1972zz, Peebles:1970ag}, has been
strongly ruled out in $\Lambda$CDM (suffering Hubble tension) at
8.4$\sigma$ \cite{Akrami:2018odb}. However,
Refs. \cite{Benetti:2017gvm,Benetti:2017juy,DiValentino:2018zjj}
point out the possibility of fully ruling out $n_s=1$ is actually
connected with the solution to the Hubble tension by noticing
qualitatively some possible correlation between a larger $n_s$ and
a larger $H_0$ in $N_{\textbf{eff}}$ (and/or
$Y_{\textbf{He}}$)+$\Lambda$CDM models. According to
(\ref{ns-H0}), the novel point of our result is that if the
current $H_0$ measured locally is correct, complete resolution of
the Hubble tension seems to be pointing to a scale invariant
Harrison-Zeldovich spectrum, i.e.$n_s\simeq 1$ for $H_0\sim
73$km/s/Mpc, see also Fig.\ref{r-ns}.

In section-\ref{sec:par shift}, we identify the physical sources
behind the parameter shifts and show the corresponding scaling
relations, which are then confronted with the MCMC results of
early resolution models in section-\ref{sec:num}. We conclude our
results in section-\ref{sec:conclusion}.

\section{Parameter shift}\label{sec:par shift}

In this section we identify the physics relevant to the
shifts of $\omega_{cdm}$, $\omega_{b}$, and $n_s$ in the early
resolutions (refer to section-\ref{sec:intro} for our definition
of ``early resolution"). Parameters in early resolutions are
expected to shift by $\lesssim15\%$ to fully resolve the Hubble
tension; thus it is sufficient to work around the
fiducial $\Lambda$CDM bestfit point. Regarding CMB
power spectra, we look for cosmological parameter shifts in
early resolutions that restore the shapes of spectra to those in
the fiducial model.

\subsection{Shift in $\omega_{cdm}$}\label{subsec:omegam}

The reason behind the $\omega_{cdm}$ shift has been presented in
Ref.\cite{Ye:2020oix}, see also \cite{Pogosian:2020ded} and the
result is $\omega_{cdm}H_0^{-2}\sim const.$ (applicable to any
early resolutions compatible with CMB and BAO data
e.g.\cite{Ye:2020oix}), or equivalently $\Omega_{cdm}\sim const.$.
Thus considering $\theta_s^*=r_s^*/D_A^*\sim const.$ and
$r_s^*H_0\sim const.$, we have
\begin{gather}
\frac{\delta H_0}{H_0}\simeq -\frac{\delta D_A}{D_A}\sim
0.5\frac{\delta \omega_{cdm}}{\omega_{cdm}}.\label{omegam-H}
\end{gather}

\subsection{Shift in $n_s$}\label{subsec:ns}
In this subsection we show how $n_s$ is shifted in response
to the change in $\omega_{b}$.

Suppose $\omega_{b}$ has some fractional deviation from its
$\Lambda$CDM value, it changes the effectiveness of baryon drag,
particularly the relative heights between even and odd TT acoustic
peaks. Baryon drag is affected by $\Psi_*$, the Newtonian
potential near last scattering, as well, but $\Psi_*$ is also
constrained independently by the (integrated) SW effect(s). Thus
$\Psi_*$ at the last-scattering surface is preserved in the early
resolutions. The peak height (PH) of the first two (1 and 2) TT
peaks numerically respond to $\omega_{b}$ according to
\begin{equation}
    \frac{\delta PH_1}{PH_1}\simeq0.3\frac{\delta\omega_b}{\omega_b},\qquad \frac{\delta PH_2}{PH_2}\simeq-0.4\frac{\delta\omega_b}{\omega_b}.
\label{PH}\end{equation} Data fitting will pin the pivot
$C_l^{TT}(k=k_{pivot})$, which is close to the second peak, to the
observed value by adjusting $A_s$ and $\tau_{reion}$ \footnote{The
overall amplitude at $l\gtrsim1500$ is suppressed by the streaming
of extra non-tightly coupled degree of freedom (e.g. $\Delta
N_{eff}$ or scalar field) in the early resolutions, so that the
increment in $A_se^{-2\tau_{reion}}$ does not spoil the fit at
high $l$.}, resulting in a cumulative excess of power in the first
peak $\frac{\delta
PH_1}{PH_1}\simeq0.7\frac{\delta\omega_b}{\omega_b}$. This is
compensated by $n_s$ according to
\begin{equation}
    \left(1+0.7\frac{\delta\omega_b}{\omega_b}\right)\left(\frac{k_1}{k_{pivot}}\right)^{\delta n_s}\sim1
\end{equation}
where $k_1\sim0.021\text{Mpc}^{-1}$ corresponds to the first TT
peak while $k_{pivot}=0.05\text{Mpc}^{-1}$ is the pivot scale.
This suggests
\begin{equation}\label{ns shift}
    \delta n_s \sim 0.8\frac{\delta \omega_b}{\omega_b}.
\end{equation}
\subsection{Shift in $\omega_b$}\label{subsec:omegab}
The damping angular scale $l_D\sim k_D D_A$ is fixed by CMB
observation \cite{Hu:1995kot}, thus $k_D$ must respond to the
fractional change in $D_A$ according to
\begin{equation}\label{kd}
    \frac{\delta k_D}{k_D}\simeq-\frac{\delta D_A}{D_A}.
\end{equation}

To have an insight into the sensitivity of $k_D$ to the background
evolution brought by the new physics shortly before recombination,
we look at a simple example. Consider new physics (e.g.dark
radiation, EDE) excited at $z_c>z_*$, which can be approximated as
a fluid with $p=w\rho$ ($w>1/3$ so it redshifts faster than radiation) at the background level, we have
\begin{equation}\label{kd int}
    \begin{aligned}
    k_D^{-2}&=\int_0^{\eta_*}\frac{d\tilde{\eta}}{6(1+R)n_e\sigma_Ta(\tilde{\eta})}\left[\frac{R^2}{1+R}+\frac{8}{9}\right]\\&\simeq k_D^{-2}(z_{c})+(27\lambda\omega_{b}
    \sigma_T/4)^{-1}\int_{a_{c}}^{a_*}\left[\omega_r(a/a_0)^{-4}+\omega_m(a/a_0)^{-3}+f_c(a/a_{c})^{-3(w+1)}\right]^{-1/2}da.
    \end{aligned}
\end{equation}
where $R\equiv3\rho_b/4\rho_\gamma$ is the baryon-to-photon energy
ratio, and $f_c$ is the energy fraction of new physics at $z_c$. $n_e\simeq\lambda\omega_{b}(a/a_0)^{-3}$, $\lambda$ being a dimensionful proportional coefficient, and $\sigma_T$ are the free electron density and Thomson cross-section respectively.
We approximately have $R=0$ since $z_c>z_*$. We set
$a_c=a_0 \omega_r/\omega_m\simeq a_{eq}$ (the matter-radiation
equality point), and expand around the $\Lambda$CDM model
\begin{equation}\label{EDE kd}
    \left|\frac{\delta k_D}{k_D}\right|\lesssim f_c\frac{\int_{1}^{y_*}y^{6-3(w+1)}(1+y)^{-3/2}dy}{\int_{1}^{y_*}y^2(1+y)^{-1/2}dy}\ll f_c
\end{equation}
where $y_*=a_*/a_c$, which shows that $k_D$ is insensitive to the
background evolution brought by the new physics before
recombination. Physically, eq.\eqref{EDE kd} represents the fact
that, whatever the new physics, its energy density redshifts fast
enough that it is negligible on the last scattering surface. The
major contribution to the integration determining $k_D$,
eq.\eqref{kd int}, comes from the last scattering surface thus the
background modification induced by the new physics has negligible
effect on $k_D$.  This suggests that the shift in $k_D$ required
by Eq.\eqref{kd} is essentially encoded in shifts of cosmological
parameters. According to $k_D\propto
\omega_b^{1/2}\omega_{cdm}^{1/4}$ (within $\Lambda$CDM) and
Eq.\eqref{omegam-H}, we get
\begin{equation}\label{omega_b tmp}
    \frac{\delta\omega_b}{\omega_b}\simeq -\frac{1}{2}\frac{\delta\omega_{cdm}}{\omega_{cdm}}-2\frac{\delta D_A}{D_A}\sim -\frac{\delta D_A}{D_A}.
\end{equation}
In $\Lambda$CDM, both the sound horizon $r_s$ (corresponding to angular scale $l\sim200$) and the damping scale $k_D$ (important for the damping tail $l>1500$) are tuned by one single parameter $\omega_{b}$. Early resolutions break this correlation by introducing new physics before recombination which only prominently affects the larger scale, i.e. $r_s$. Actually, the increment in $\omega_{b}$ will be less than
Eq.\eqref{omega_b tmp}, since compared with the $\Lambda$CDM model
some extra damping is needed to compensate for the excess power at
high $l$ brought by a larger $n_s$. However, since the high $l$
CMB data is not as precise as the first few acoustic peaks, it is
difficult to speculate the corresponding effects in an analytical
way. To this end, we marginalize over this effect with the
parameter $0<\alpha<1$ and rewrite (\ref{omega_b tmp}) as
\begin{equation}\label{omega_b}
    \frac{\delta\omega_b}{\omega_b}\sim-(1-\alpha)\frac{\delta D_A}{D_A}.
\end{equation}



\section{$n_s$-$H_0$ scaling relation and MCMC results}\label{sec:num}

We confront the scaling relations shown in
section-\ref{sec:par shift} with the MCMC results. As concrete
examples of early resolution models, we limit ourself to the EDE.
The EDE models we consider are the $n=3$ Axion model
$V(\phi)=V_0(1-\cos(\phi/f))^3$
\cite{Poulin:2018cxd,Smith:2019ihp}, the $n=2$ Rock `n' model
$V=V_0(\phi/M_p)^4$ \cite{Agrawal:2019lmo} (called $\phi^4$ for
simplicity) and the AdS-EDE model with fixed AdS depth, see
\cite{Ye:2020btb} for details. In addition to the six $\Lambda$CDM
cosmological parameters $\{\omega_{b},\omega_{cdm},H_0,
\ln10^{10}A_s,n_s,\tau_{reion}\}$, all EDE models have two
additional MCMC parameters $\{\ln(1+z_c),f_{ede}\}$, with $z_c$
being the redshift at which the field $\phi$ starts rolling and
$f_{ede}$ the energy fraction of EDE at $z_c$. The Axion
model varies yet one more MCMC parameter $\Theta_i$, the
initial position of the scalar field, see \cite{Poulin:2018cxd}
for details.

According to Eqs.\eqref{omegam-H}, \eqref{ns shift} and
\eqref{omega_b}, the shift of parameters
$\{\omega_{cdm},\omega_b,n_s\}$ can be straightly related to
$f_{ede}$. Generally, all components (baryon, dark matter,
radiation\footnote{The radiation energy density is fixed by the
$T_{0,FIRAS}$ \cite{Fixsen:1996nj,Fixsen:2009ug}, which is
compatible with EDE \cite{Ye:2020oix}.} and early dark energy)
contribute to $r_s^*$. Assuming the energy injection near
matter-radiation equality $z_c\approx z_{eq}$, which is valid for
almost all EDE models, we numerically evaluate the response of $r_s$ to
$f_{ede}$, $\omega_{cdm}$ and $\omega_{b}$ around the $\Lambda$CDM
bestfit ($f_{ede}=0$)
\begin{equation}\label{rs-ede}
-\frac{\delta r_s}{r_s}\simeq 0.3f_{ede}+0.2\frac{\delta
\omega_{cdm}}{\omega_{cdm}}+0.1\frac{\delta
\omega_{b}}{\omega_{b}}.
\end{equation}
Compatibility with \eqref{omegam-H} implies $\frac{\delta
\omega_{cdm}}{\omega_{cdm}}\simeq f_{ede}+0.33\frac{\delta
\omega_{b}}{\omega_{b}}$. This is equivalent to adjusting $f_{ede}$
and $\omega_{cdm}$ such that near recombination
$\Phi^{EDE}(l)\simeq\Phi^{LCDM}(l)$ up to data uncertainty for the
first few peaks. Thus we have
\begin{gather}
\frac{\delta\omega_{b}}{\omega_{b}}\sim0.6f_{ede}(1-1.2\alpha),\label{omegab_ede}\\
\delta n_s\sim0.5f_{ede}(1-1.2\alpha),\quad \frac{\delta
H_0}{H_0}\simeq 0.5\frac{\delta\omega_{cdm}}{\omega_{cdm}}\sim
0.6f_{ede}(1-0.2\alpha).\label{ede parshift}
\end{gather}
Thus $H_0$ is lifted proportionally to $f_{ede}>0$. However, the
cost of making EDE still fit CMB, BAO and light curve
observations as well as $\Lambda$CDM does (as is confirmed with
the MCMC analysis) is that the relevant parameters must be shifted
($\sim f_{ede}$).

To clearly see the effect of the Hubble tension on the parameters
$n_s$ and $r$ of the primordial Universe, where the
tensor-to-scalar ratio $r=A_T/A_s$
($k_{pivot}=0.05\text{Mpc}^{-1}$), we use Planck low-$l$ EEBB
and Keck Array/BICEP 2015 data \cite{Array:2015xqh}, and set
recent SH0ES result $H_0=74.03\pm 1.42$km/s/Mpc
\cite{Riess:2019cxk} (R19) as a Gaussian prior. In addition, our
datasets consist of the Planck18 high-$l$ TTTEEE and low-$l$ TT
likelihoods as well as Planck lensing \cite{Aghanim:2018eyx}, the
BOSS DR12 \cite{Alam:2016hwk} with its full covariant matrix for
BAO as well as the 6dFGS \cite{Beutler:2011hx} and MGS of SDSS
\cite{Ross:2014qpa} for low-$z$ BAO, and the Pantheon data
\cite{Scolnic:2017caz}.

It should be underlined that the fiducial model we consider is
$\Lambda$CDM with its six cosmological parameters and the MCMC
results for $\Lambda$CDM depend on dataset. Our results in
section-\ref{sec:par shift} are based on Eq.(\ref{omegam-H}),
which suggests that the corresponding MCMC dataset must include
CMB and BAO data at least \cite{Ye:2020oix}.

\begin{figure}
\centering \subfigure[$\omega_{cdm}-H_0$] { \label{MID-mh}
    \includegraphics[width=0.48\linewidth]{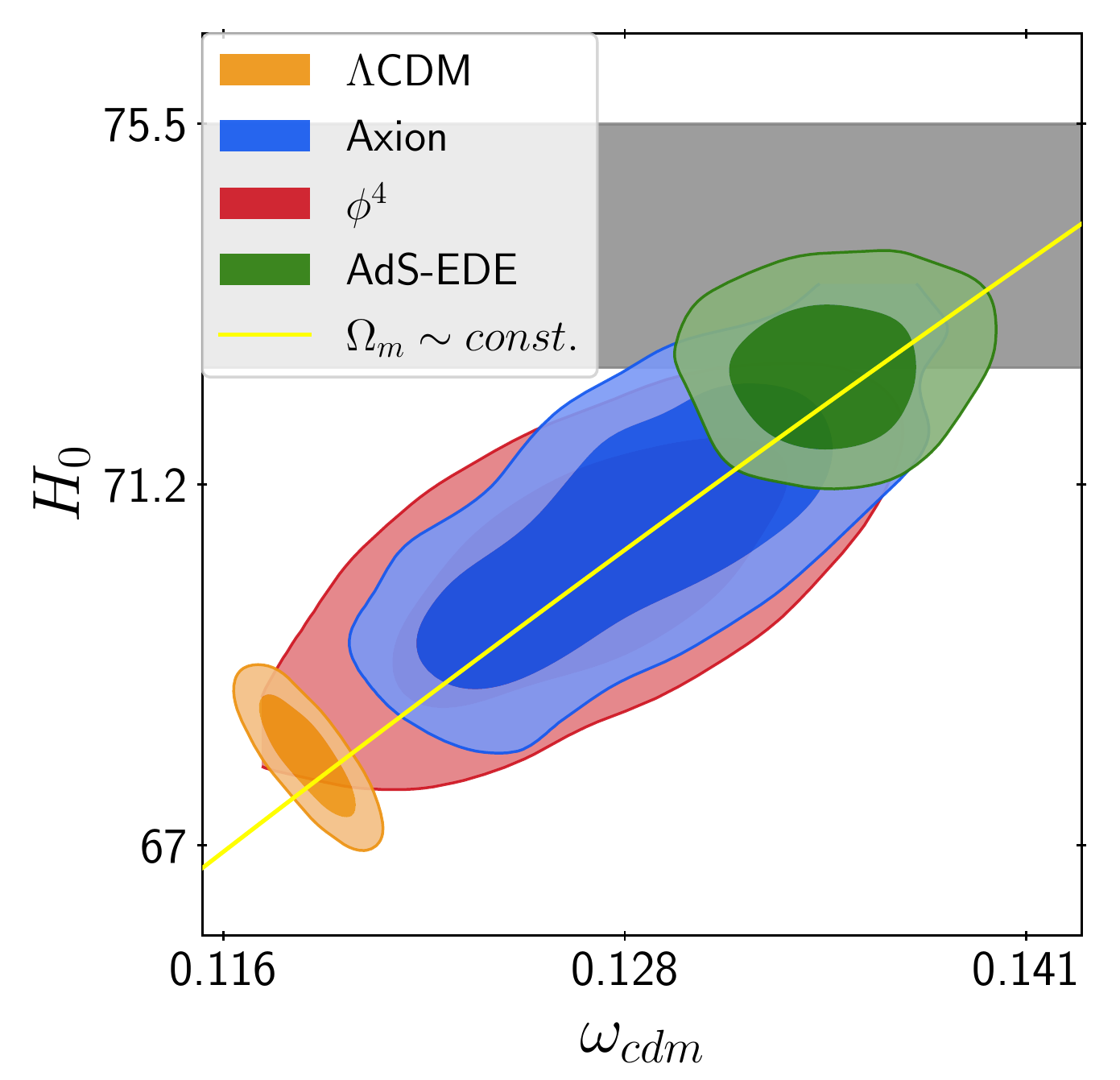}}
\subfigure[$\omega_{b}-n_s$] { \label{MID-nb}
    \includegraphics[width=0.48\linewidth]{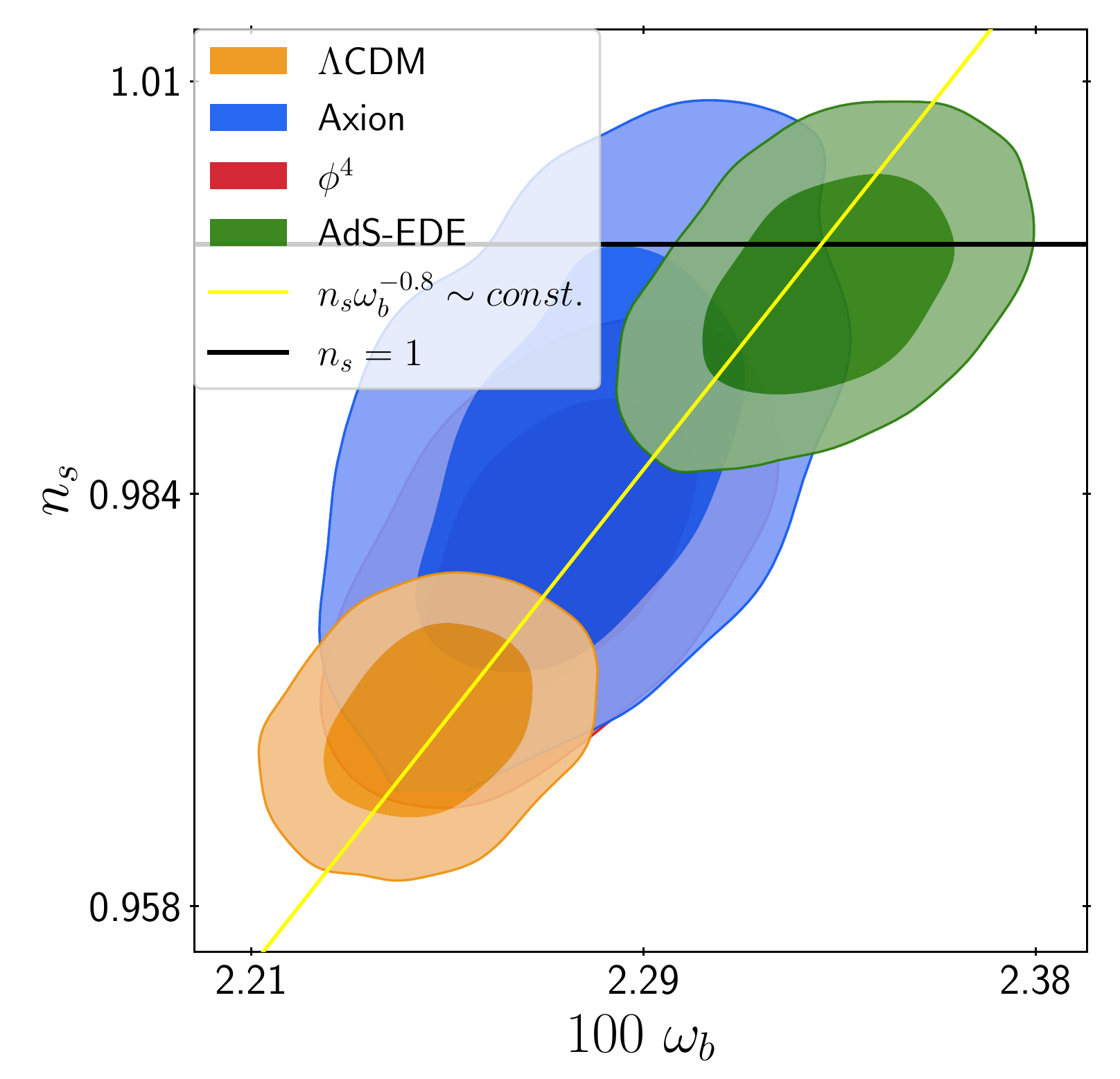}}
\caption{Analytic approximations \eqref{omegam-H} and \eqref{ns
shift} confronted with MCMC 68\% and 95\% contours for EDE models.
\textit{Left panel}: $\omega_{cdm}$ vs. $H_0$. Yellow line plots
Eq.\eqref{omegam-H}. \textit{Right panel}: $\omega_b$ vs. $n_s$.
Yellow line plots Eq.\eqref{ns shift}.} \label{MID}
\end{figure}

We modified the Montepython-3.3
\cite{Audren:2012wb,Brinckmann:2018cvx} and CLASS
\cite{Lesgourgues:2011re,Blas:2011rf} codes to perform the MCMC
analysis. Table-\ref{ede tab} presents the MCMC results for
$\Lambda$CDM and EDE models (Axion, $\phi^4$ and AdS-EDE), see
also the corresponding $H_0$-$n_s$ and $r$-$n_s$ contours in
Figs.\ref{H0-ns} and \ref{r-ns}, respectively. As expected, all
early resolution models fit to CMB and BAO as well as $\Lambda$CDM
does, see Table.\ref{chi2} for the bestfit $\chi^2$ per
experiment. The existence of the AdS region in AdS-EDE actually
sets a physical lower bound on the EDE energy fraction $f_{ede}$,
because the field would fail to climb out of the AdS region if
$f_{ede}$ is too small. However, as is clear in Fig.\ref{H0-ns},
\ref{r-ns}, \ref{MID} and Table.\ref{chi2}, the MCMC chain is not
hard capped by this bound and converges around the bestfit point
well. The upper bound on $r$ in the AdS-EDE model is slightly smaller
than that in other models, see Fig.\ref{r-ns} and Table-\ref{ede
tab}.

\begin{table}
    \begin{tabular}{|c|c|c|c|c|c|c||c|}
        \hline
        &$f_{ede}$&$100\omega_b$&$\alpha$&$H_0$&$\omega_{cdm}$&$n_s$&$r$\\
        \hline
        $\Lambda$CDM&-&2.246&-&68.1&0.1184&0.969&$<0.0636$\\
        \hline
        $\phi^4$&0.070&2.274&0.59&70.3(70.6)&0.1271(0.127)&0.980(0.979)&$<0.0603$\\
        \hline
        Axion&0.094&2.295&0.51&70.9(71.5)&0.1295(0.13)&0.987(0.987)&$<0.066$\\
        \hline
        AdS-EDE&0.115&2.336&0.35&72.6(72.5)&0.1346(0.134)&0.997(1)&$<0.0574$\\
        \hline
    \end{tabular}
    \caption{The mean values of parameters in corresponding models and
        the 95\% upper bounds on the tensor-to-scalar ratio $r$. In the
        parenthesis are the analytic estimations made by Eq.\eqref{ede
            parshift}, which are consistent with the MCMC results. }
    \label{ede tab}
\end{table}

Actual numerical results of parameter shifts are plotted in
Fig.\ref{MID}, which are consistent with Eqs.\eqref{omegam-H} and
\eqref{ns shift}. Though the detailed shape of CMB spectra is
affected by all cosmological parameters in complicated ways, the
analytic approximations Eqs.\eqref{omegam-H} and \eqref{ns shift}
have clearly captured the common pattern of parameter shifts.

As representative early resolutions models, all EDE models in
Figs.\ref{H0-ns} and \ref{r-ns} show $n_s\gtrsim 0.98$, which is
actually a ``\textit{universal}" prediction of early resolutions
of the Hubble tension, in particular the AdS-EDE model allows
$n_s=1$ at 1$\sigma$ region. To see this, Eqs.\eqref{omegam-H},
\eqref{ns shift} and \eqref{omega_b} in combination relates $n_s$
with $H_0$
\begin{equation}\label{ns-H}
    \delta n_s\sim 0.8(1-\alpha)\frac{\delta H_0}{H_0}.
\end{equation}
Here, $\alpha$ can be set by Eq.(\ref{omegab_ede}) with MCMC
results of $f_{ede}$ and $\omega_b$, see Table-\ref{ede tab}.
Generally, we have $0.4\lesssim \alpha\lesssim 0.6$. According to
(\ref{ns-H}), we approximately get (\ref{ns-H0}).

The argument in section-\ref{sec:par shift} is valid for early
resolutions, as defined in section-\ref{sec:intro}, with rather
weak dependence on the specific physical model. According to
(\ref{ns-H}), we have $n_s\simeq 1$ for $H_0\sim 73$km/s/Mpc. Thus
it is intriguing to speculate that, contrary to $\Lambda$CDM,
early resolutions of the Hubble tension might play well with
$n_s\simeq 1$.

\begin{table}
    \begin{tabular}{|c|c|c|c|c|}
        \hline
        Experiment&$\Lambda$CDM&$\phi^4$&Axion&AdS-EDE\\
        \hline
        Planck high $l$ TTTEEE&2347.1&2347.1&2346.4&2349.9 \\
        \hline
        Planck low $l$ TT&22.7&21.9&20.8&21.1 \\
        \hline
        Planck low $l$ EEBB&785.9&785.7&785.9&785.5 \\
        \hline
        Planck lensing&10.2&9.7&10.9&10.4 \\
        \hline
        BK15&736.5&737.8&738.8&734.1 \\
        \hline
        BOSS DR12&3.4&4.3&3.9&3.5 \\
        \hline
        BAO low-$z$&1.9&1.3&1.4&1.8 \\
        \hline
        Pantheon&1026.9&1027.3&1027.2&1026.9 \\
        \hline
        $H_0$ prior&15.6&8.3&6.2&1.3 \\
        \hline
    \end{tabular}
    \caption{Bestfit $\chi^2$ per experiment.}
    \label{chi2}
\end{table}

\section{Conclusion}\label{sec:conclusion}

The early resolutions of the Hubble tension, such as EDE, have
been found to fit with CMB, BAO and light curve observations as
well as $\Lambda$CDM does, however, the cost of compensating for the
impact of new physics before recombination is that the values of
relevant parameters must be shifted. We have identified the major
physical source behind the parameter shifts. The patterns of
parameter shifts are represented by a set of linear response
equations \eqref{omegam-H}, \eqref{ns shift}, \eqref{omega_b} and
\eqref{ns-H}, which are confirmed by performing the MCMC analysis
with Planck2018+BAO+Pantheon+R19 dataset as well as Keck
Array/BICEP data. Our results are common to early resolutions
defined in section-\ref{sec:intro}, which not only bring new
insight into the physics behind the early resolutions, but also
highlight the significance of other model-independent probes of
$n_s$, $\omega_{cdm}$, etc. for falsifying early resolutions,
e.g.\cite{Lucca:2020fgp, Krishnan:2020vaf, Bernal:2021yli, Fanizza:2021tuh}.

Specially, the shift of $n_s$ with respect to $\delta H$ implies
that data (Planck2018+Keck Array/BICEP+BAO+Pantheon+R19) allow
for much larger $n_s$ in the early resolutions than in
$\Lambda$CDM. Interestingly, if the current local measurement
$H_0\sim73$km/s/Mpc is correct, it seems to favor an $n_s=1$
cosmology in the early resolutions, which has been strongly ruled
out in $\Lambda$CDM \cite{Akrami:2018odb}. This might have
profound implication to inflation and the early Universe physics.
Compared with Planck 2018 result \cite{Akrami:2018odb}, a
scale-invariant primordial spectrum together with the current
upper bound on tensor-to-scalar ratio $r\lesssim0.06$
(Fig.\ref{r-ns}) does not favor e.g. Starobinski inflation,
hilltop inflation. Theoretical implications of our observation is
yet to be explored, however, it would be expected that the
primordial universe with $n_s=1$ might be radically different from
the popular paradigm of slow-roll inflation.

Another potentially interesting point is the degeneracy between
$n_s$ and diffusion damping, as mentioned in
section-\ref{subsec:omegab}. Precise CMB power spectra
observations at $l\gtrsim2000$ might help break this degeneracy.
See, for example,
Refs.\cite{Chudaykin:2020acu,Chudaykin:2020igl,Lin:2020jcb} for
current proceedings with SPT \cite{Henning:2017nuy} or ACT
\cite{Choi:2020ccd}.

\paragraph*{Acknowledgments}
YSP is supported by the National Natural Science Foundation of China Grants Nos.12075246, 11690021.
BH is supported by the National Natural Science Foundation of China Grants No. 11973016.
To produce
results of the $\phi^4$ and Axion model we make use of the
publicly available codes class\_rnr
(\url{https://github.com/franyancr/class_rnr}) and AxiCLASS
(\url{https://github.com/PoulinV/AxiCLASS}).
\bibliography{ref}

\begin{thebibliography}{79}
\expandafter\ifx\csname natexlab\endcsname\relax\def\natexlab#1{#1}\fi
\expandafter\ifx\csname bibnamefont\endcsname\relax
  \def\bibnamefont#1{#1}\fi
\expandafter\ifx\csname bibfnamefont\endcsname\relax
  \def\bibfnamefont#1{#1}\fi
\expandafter\ifx\csname citenamefont\endcsname\relax
  \def\citenamefont#1{#1}\fi
\expandafter\ifx\csname url\endcsname\relax
  \def\url#1{\texttt{#1}}\fi
\expandafter\ifx\csname urlprefix\endcsname\relax\def\urlprefix{URL }\fi
\providecommand{\bibinfo}[2]{#2}
\providecommand{\eprint}[2][]{\url{#2}}

\bibitem[{\citenamefont{Aghanim et~al.}(2020)}]{Aghanim:2018eyx}
\bibinfo{author}{\bibfnamefont{N.}~\bibnamefont{Aghanim}} \bibnamefont{et~al.}
  (\bibinfo{collaboration}{Planck}), \bibinfo{journal}{Astron. Astrophys.}
  \textbf{\bibinfo{volume}{641}}, \bibinfo{pages}{A6} (\bibinfo{year}{2020}),
  \eprint{1807.06209}.

\bibitem[{\citenamefont{Verde et~al.}(2019)\citenamefont{Verde, Treu, and
  Riess}}]{Verde:2019ivm}
\bibinfo{author}{\bibfnamefont{L.}~\bibnamefont{Verde}},
  \bibinfo{author}{\bibfnamefont{T.}~\bibnamefont{Treu}}, \bibnamefont{and}
  \bibinfo{author}{\bibfnamefont{A.}~\bibnamefont{Riess}},
  \bibinfo{journal}{Nature Astron.} \textbf{\bibinfo{volume}{3}},
  \bibinfo{pages}{891} (\bibinfo{year}{2019}), \eprint{1907.10625}.

\bibitem[{\citenamefont{Riess}(2019)}]{Riess:2020sih}
\bibinfo{author}{\bibfnamefont{A.~G.} \bibnamefont{Riess}},
  \bibinfo{journal}{Nature Rev. Phys.} \textbf{\bibinfo{volume}{2}},
  \bibinfo{pages}{10} (\bibinfo{year}{2019}), \eprint{2001.03624}.

\bibitem[{\citenamefont{Bernal et~al.}(2016)\citenamefont{Bernal, Verde, and
  Riess}}]{Bernal:2016gxb}
\bibinfo{author}{\bibfnamefont{J.~L.} \bibnamefont{Bernal}},
  \bibinfo{author}{\bibfnamefont{L.}~\bibnamefont{Verde}}, \bibnamefont{and}
  \bibinfo{author}{\bibfnamefont{A.~G.} \bibnamefont{Riess}},
  \bibinfo{journal}{JCAP} \textbf{\bibinfo{volume}{10}}, \bibinfo{pages}{019}
  (\bibinfo{year}{2016}), \eprint{1607.05617}.

\bibitem[{\citenamefont{Feeney et~al.}(2018)\citenamefont{Feeney, Mortlock, and
  Dalmasso}}]{Feeney:2017sgx}
\bibinfo{author}{\bibfnamefont{S.~M.} \bibnamefont{Feeney}},
  \bibinfo{author}{\bibfnamefont{D.~J.} \bibnamefont{Mortlock}},
  \bibnamefont{and} \bibinfo{author}{\bibfnamefont{N.}~\bibnamefont{Dalmasso}},
  \bibinfo{journal}{Mon. Not. Roy. Astron. Soc.}
  \textbf{\bibinfo{volume}{476}}, \bibinfo{pages}{3861} (\bibinfo{year}{2018}),
  \eprint{1707.00007}.

\bibitem[{\citenamefont{Aylor et~al.}(2019)\citenamefont{Aylor, Joy, Knox,
  Millea, Raghunathan, and Wu}}]{Aylor:2018drw}
\bibinfo{author}{\bibfnamefont{K.}~\bibnamefont{Aylor}},
  \bibinfo{author}{\bibfnamefont{M.}~\bibnamefont{Joy}},
  \bibinfo{author}{\bibfnamefont{L.}~\bibnamefont{Knox}},
  \bibinfo{author}{\bibfnamefont{M.}~\bibnamefont{Millea}},
  \bibinfo{author}{\bibfnamefont{S.}~\bibnamefont{Raghunathan}},
  \bibnamefont{and} \bibinfo{author}{\bibfnamefont{W.~K.} \bibnamefont{Wu}},
  \bibinfo{journal}{Astrophys. J.} \textbf{\bibinfo{volume}{874}},
  \bibinfo{pages}{4} (\bibinfo{year}{2019}), \eprint{1811.00537}.

\bibitem[{\citenamefont{Knox and Millea}(2020)}]{Knox:2019rjx}
\bibinfo{author}{\bibfnamefont{L.}~\bibnamefont{Knox}} \bibnamefont{and}
  \bibinfo{author}{\bibfnamefont{M.}~\bibnamefont{Millea}},
  \bibinfo{journal}{Phys. Rev. D} \textbf{\bibinfo{volume}{101}},
  \bibinfo{pages}{043533} (\bibinfo{year}{2020}), \eprint{1908.03663}.

\bibitem[{\citenamefont{Di~Valentino
  et~al.}(2020{\natexlab{a}})}]{DiValentino:2020zio}
\bibinfo{author}{\bibfnamefont{E.}~\bibnamefont{Di~Valentino}}
  \bibnamefont{et~al.} (\bibinfo{year}{2020}{\natexlab{a}}),
  \eprint{2008.11284}.

\bibitem[{\citenamefont{Lyu et~al.}(2020)\citenamefont{Lyu, Haridasu, Viel, and
  Xia}}]{Lyu:2020lwm}
\bibinfo{author}{\bibfnamefont{M.-Z.} \bibnamefont{Lyu}},
  \bibinfo{author}{\bibfnamefont{B.~S.} \bibnamefont{Haridasu}},
  \bibinfo{author}{\bibfnamefont{M.}~\bibnamefont{Viel}}, \bibnamefont{and}
  \bibinfo{author}{\bibfnamefont{J.-Q.} \bibnamefont{Xia}},
  \bibinfo{journal}{Astrophys. J.} \textbf{\bibinfo{volume}{900}},
  \bibinfo{pages}{160} (\bibinfo{year}{2020}), \eprint{2001.08713}.

\bibitem[{\citenamefont{Haridasu et~al.}(2020)\citenamefont{Haridasu, Viel, and
  Vittorio}}]{Haridasu:2020pms}
\bibinfo{author}{\bibfnamefont{B.~S.} \bibnamefont{Haridasu}},
  \bibinfo{author}{\bibfnamefont{M.}~\bibnamefont{Viel}}, \bibnamefont{and}
  \bibinfo{author}{\bibfnamefont{N.}~\bibnamefont{Vittorio}}
  (\bibinfo{year}{2020}), \eprint{2012.10324}.

\bibitem[{\citenamefont{Di~Valentino et~al.}(2021)\citenamefont{Di~Valentino,
  Mena, Pan, Visinelli, Yang, Melchiorri, Mota, Riess, and
  Silk}}]{DiValentino:2021izs}
\bibinfo{author}{\bibfnamefont{E.}~\bibnamefont{Di~Valentino}},
  \bibinfo{author}{\bibfnamefont{O.}~\bibnamefont{Mena}},
  \bibinfo{author}{\bibfnamefont{S.}~\bibnamefont{Pan}},
  \bibinfo{author}{\bibfnamefont{L.}~\bibnamefont{Visinelli}},
  \bibinfo{author}{\bibfnamefont{W.}~\bibnamefont{Yang}},
  \bibinfo{author}{\bibfnamefont{A.}~\bibnamefont{Melchiorri}},
  \bibinfo{author}{\bibfnamefont{D.~F.} \bibnamefont{Mota}},
  \bibinfo{author}{\bibfnamefont{A.~G.} \bibnamefont{Riess}}, \bibnamefont{and}
  \bibinfo{author}{\bibfnamefont{J.}~\bibnamefont{Silk}}
  (\bibinfo{year}{2021}), \eprint{2103.01183}.

\bibitem[{\citenamefont{Arest\'e~Sal\'o
  et~al.}(2021{\natexlab{a}})\citenamefont{Arest\'e~Sal\'o, Benisty,
  Guendelman, and de~Haro}}]{AresteSalo:2021wgb}
\bibinfo{author}{\bibfnamefont{L.}~\bibnamefont{Arest\'e~Sal\'o}},
  \bibinfo{author}{\bibfnamefont{D.}~\bibnamefont{Benisty}},
  \bibinfo{author}{\bibfnamefont{E.~I.} \bibnamefont{Guendelman}},
  \bibnamefont{and} \bibinfo{author}{\bibfnamefont{J.}~\bibnamefont{de~Haro}}
  (\bibinfo{year}{2021}{\natexlab{a}}), \eprint{2103.07892}.

\bibitem[{\citenamefont{Arest\'e~Sal\'o
  et~al.}(2021{\natexlab{b}})\citenamefont{Arest\'e~Sal\'o, Benisty,
  Guendelman, and Haro}}]{AresteSalo:2021lmp}
\bibinfo{author}{\bibfnamefont{L.}~\bibnamefont{Arest\'e~Sal\'o}},
  \bibinfo{author}{\bibfnamefont{D.}~\bibnamefont{Benisty}},
  \bibinfo{author}{\bibfnamefont{E.~I.} \bibnamefont{Guendelman}},
  \bibnamefont{and} \bibinfo{author}{\bibfnamefont{J.~d.} \bibnamefont{Haro}}
  (\bibinfo{year}{2021}{\natexlab{b}}), \eprint{2102.09514}.

\bibitem[{\citenamefont{Dainotti et~al.}(2021)\citenamefont{Dainotti,
  De~Simone, Schiavone, Montani, Rinaldi, and Lambiase}}]{Dainotti:2021pqg}
\bibinfo{author}{\bibfnamefont{M.~G.} \bibnamefont{Dainotti}},
  \bibinfo{author}{\bibfnamefont{B.}~\bibnamefont{De~Simone}},
  \bibinfo{author}{\bibfnamefont{T.}~\bibnamefont{Schiavone}},
  \bibinfo{author}{\bibfnamefont{G.}~\bibnamefont{Montani}},
  \bibinfo{author}{\bibfnamefont{E.}~\bibnamefont{Rinaldi}}, \bibnamefont{and}
  \bibinfo{author}{\bibfnamefont{G.}~\bibnamefont{Lambiase}}
  (\bibinfo{year}{2021}), \eprint{2103.02117}.

\bibitem[{\citenamefont{Poulin et~al.}(2019)\citenamefont{Poulin, Smith,
  Karwal, and Kamionkowski}}]{Poulin:2018cxd}
\bibinfo{author}{\bibfnamefont{V.}~\bibnamefont{Poulin}},
  \bibinfo{author}{\bibfnamefont{T.~L.} \bibnamefont{Smith}},
  \bibinfo{author}{\bibfnamefont{T.}~\bibnamefont{Karwal}}, \bibnamefont{and}
  \bibinfo{author}{\bibfnamefont{M.}~\bibnamefont{Kamionkowski}},
  \bibinfo{journal}{Phys. Rev. Lett.} \textbf{\bibinfo{volume}{122}},
  \bibinfo{pages}{221301} (\bibinfo{year}{2019}), \eprint{1811.04083}.

\bibitem[{\citenamefont{Agrawal et~al.}(2019)\citenamefont{Agrawal, Cyr-Racine,
  Pinner, and Randall}}]{Agrawal:2019lmo}
\bibinfo{author}{\bibfnamefont{P.}~\bibnamefont{Agrawal}},
  \bibinfo{author}{\bibfnamefont{F.-Y.} \bibnamefont{Cyr-Racine}},
  \bibinfo{author}{\bibfnamefont{D.}~\bibnamefont{Pinner}}, \bibnamefont{and}
  \bibinfo{author}{\bibfnamefont{L.}~\bibnamefont{Randall}}
  (\bibinfo{year}{2019}), \eprint{1904.01016}.

\bibitem[{\citenamefont{Lin et~al.}(2019)\citenamefont{Lin, Benevento, Hu, and
  Raveri}}]{Lin:2019qug}
\bibinfo{author}{\bibfnamefont{M.-X.} \bibnamefont{Lin}},
  \bibinfo{author}{\bibfnamefont{G.}~\bibnamefont{Benevento}},
  \bibinfo{author}{\bibfnamefont{W.}~\bibnamefont{Hu}}, \bibnamefont{and}
  \bibinfo{author}{\bibfnamefont{M.}~\bibnamefont{Raveri}},
  \bibinfo{journal}{Phys. Rev. D} \textbf{\bibinfo{volume}{100}},
  \bibinfo{pages}{063542} (\bibinfo{year}{2019}), \eprint{1905.12618}.

\bibitem[{\citenamefont{Ye and Piao}(2020{\natexlab{a}})}]{Ye:2020btb}
\bibinfo{author}{\bibfnamefont{G.}~\bibnamefont{Ye}} \bibnamefont{and}
  \bibinfo{author}{\bibfnamefont{Y.-S.} \bibnamefont{Piao}},
  \bibinfo{journal}{Phys. Rev. D} \textbf{\bibinfo{volume}{101}},
  \bibinfo{pages}{083507} (\bibinfo{year}{2020}{\natexlab{a}}),
  \eprint{2001.02451}.

\bibitem[{\citenamefont{Alexander and McDonough}(2019)}]{Alexander:2019rsc}
\bibinfo{author}{\bibfnamefont{S.}~\bibnamefont{Alexander}} \bibnamefont{and}
  \bibinfo{author}{\bibfnamefont{E.}~\bibnamefont{McDonough}},
  \bibinfo{journal}{Phys. Lett. B} \textbf{\bibinfo{volume}{797}},
  \bibinfo{pages}{134830} (\bibinfo{year}{2019}), \eprint{1904.08912}.

\bibitem[{\citenamefont{Smith et~al.}(2020)\citenamefont{Smith, Poulin, and
  Amin}}]{Smith:2019ihp}
\bibinfo{author}{\bibfnamefont{T.~L.} \bibnamefont{Smith}},
  \bibinfo{author}{\bibfnamefont{V.}~\bibnamefont{Poulin}}, \bibnamefont{and}
  \bibinfo{author}{\bibfnamefont{M.~A.} \bibnamefont{Amin}},
  \bibinfo{journal}{Phys. Rev. D} \textbf{\bibinfo{volume}{101}},
  \bibinfo{pages}{063523} (\bibinfo{year}{2020}), \eprint{1908.06995}.

\bibitem[{\citenamefont{Niedermann and Sloth}(2019)}]{Niedermann:2019olb}
\bibinfo{author}{\bibfnamefont{F.}~\bibnamefont{Niedermann}} \bibnamefont{and}
  \bibinfo{author}{\bibfnamefont{M.~S.} \bibnamefont{Sloth}}
  (\bibinfo{year}{2019}), \eprint{1910.10739}.

\bibitem[{\citenamefont{Sakstein and Trodden}(2020)}]{Sakstein:2019fmf}
\bibinfo{author}{\bibfnamefont{J.}~\bibnamefont{Sakstein}} \bibnamefont{and}
  \bibinfo{author}{\bibfnamefont{M.}~\bibnamefont{Trodden}},
  \bibinfo{journal}{Phys. Rev. Lett.} \textbf{\bibinfo{volume}{124}},
  \bibinfo{pages}{161301} (\bibinfo{year}{2020}), \eprint{1911.11760}.

\bibitem[{\citenamefont{Chudaykin
  et~al.}(2020{\natexlab{a}})\citenamefont{Chudaykin, Gorbunov, and
  Nedelko}}]{Chudaykin:2020acu}
\bibinfo{author}{\bibfnamefont{A.}~\bibnamefont{Chudaykin}},
  \bibinfo{author}{\bibfnamefont{D.}~\bibnamefont{Gorbunov}}, \bibnamefont{and}
  \bibinfo{author}{\bibfnamefont{N.}~\bibnamefont{Nedelko}},
  \bibinfo{journal}{JCAP} \textbf{\bibinfo{volume}{08}}, \bibinfo{pages}{013}
  (\bibinfo{year}{2020}{\natexlab{a}}), \eprint{2004.13046}.

\bibitem[{\citenamefont{Ye and Piao}(2020{\natexlab{b}})}]{Ye:2020oix}
\bibinfo{author}{\bibfnamefont{G.}~\bibnamefont{Ye}} \bibnamefont{and}
  \bibinfo{author}{\bibfnamefont{Y.-S.} \bibnamefont{Piao}},
  \bibinfo{journal}{Phys. Rev. D} \textbf{\bibinfo{volume}{102}},
  \bibinfo{pages}{083523} (\bibinfo{year}{2020}{\natexlab{b}}),
  \eprint{2008.10832}.

\bibitem[{\citenamefont{Braglia
  et~al.}(2020{\natexlab{a}})\citenamefont{Braglia, Ballardini, Emond, Finelli,
  Gumrukcuoglu, Koyama, and Paoletti}}]{Braglia:2020iik}
\bibinfo{author}{\bibfnamefont{M.}~\bibnamefont{Braglia}},
  \bibinfo{author}{\bibfnamefont{M.}~\bibnamefont{Ballardini}},
  \bibinfo{author}{\bibfnamefont{W.~T.} \bibnamefont{Emond}},
  \bibinfo{author}{\bibfnamefont{F.}~\bibnamefont{Finelli}},
  \bibinfo{author}{\bibfnamefont{A.~E.} \bibnamefont{Gumrukcuoglu}},
  \bibinfo{author}{\bibfnamefont{K.}~\bibnamefont{Koyama}}, \bibnamefont{and}
  \bibinfo{author}{\bibfnamefont{D.}~\bibnamefont{Paoletti}},
  \bibinfo{journal}{Phys. Rev. D} \textbf{\bibinfo{volume}{102}},
  \bibinfo{pages}{023529} (\bibinfo{year}{2020}{\natexlab{a}}),
  \eprint{2004.11161}.

\bibitem[{\citenamefont{Lin et~al.}(2020)\citenamefont{Lin, Hu, and
  Raveri}}]{Lin:2020jcb}
\bibinfo{author}{\bibfnamefont{M.-X.} \bibnamefont{Lin}},
  \bibinfo{author}{\bibfnamefont{W.}~\bibnamefont{Hu}}, \bibnamefont{and}
  \bibinfo{author}{\bibfnamefont{M.}~\bibnamefont{Raveri}},
  \bibinfo{journal}{Phys. Rev. D} \textbf{\bibinfo{volume}{102}},
  \bibinfo{pages}{123523} (\bibinfo{year}{2020}), \eprint{2009.08974}.

\bibitem[{\citenamefont{Niedermann and Sloth}(2020)}]{Niedermann:2020dwg}
\bibinfo{author}{\bibfnamefont{F.}~\bibnamefont{Niedermann}} \bibnamefont{and}
  \bibinfo{author}{\bibfnamefont{M.~S.} \bibnamefont{Sloth}},
  \bibinfo{journal}{Phys. Rev. D} \textbf{\bibinfo{volume}{102}},
  \bibinfo{pages}{063527} (\bibinfo{year}{2020}), \eprint{2006.06686}.

\bibitem[{\citenamefont{Chudaykin
  et~al.}(2020{\natexlab{b}})\citenamefont{Chudaykin, Gorbunov, and
  Nedelko}}]{Chudaykin:2020igl}
\bibinfo{author}{\bibfnamefont{A.}~\bibnamefont{Chudaykin}},
  \bibinfo{author}{\bibfnamefont{D.}~\bibnamefont{Gorbunov}}, \bibnamefont{and}
  \bibinfo{author}{\bibfnamefont{N.}~\bibnamefont{Nedelko}}
  (\bibinfo{year}{2020}{\natexlab{b}}), \eprint{2011.04682}.

\bibitem[{\citenamefont{Fujita et~al.}(2020)\citenamefont{Fujita, Murai,
  Nakatsuka, and Tsujikawa}}]{Fujita:2020ecn}
\bibinfo{author}{\bibfnamefont{T.}~\bibnamefont{Fujita}},
  \bibinfo{author}{\bibfnamefont{K.}~\bibnamefont{Murai}},
  \bibinfo{author}{\bibfnamefont{H.}~\bibnamefont{Nakatsuka}},
  \bibnamefont{and} \bibinfo{author}{\bibfnamefont{S.}~\bibnamefont{Tsujikawa}}
  (\bibinfo{year}{2020}), \eprint{2011.11894}.

\bibitem[{\citenamefont{Seto and Toda}(2021)}]{Seto:2021xua}
\bibinfo{author}{\bibfnamefont{O.}~\bibnamefont{Seto}} \bibnamefont{and}
  \bibinfo{author}{\bibfnamefont{Y.}~\bibnamefont{Toda}}
  (\bibinfo{year}{2021}), \eprint{2101.03740}.

\bibitem[{\citenamefont{Tian and Zhu}(2021)}]{Tian:2021omz}
\bibinfo{author}{\bibfnamefont{S.~X.} \bibnamefont{Tian}} \bibnamefont{and}
  \bibinfo{author}{\bibfnamefont{Z.-H.} \bibnamefont{Zhu}},
  \bibinfo{journal}{Phys. Rev. D} \textbf{\bibinfo{volume}{103}},
  \bibinfo{pages}{043518} (\bibinfo{year}{2021}), \eprint{2102.06399}.

\bibitem[{\citenamefont{Sabla and Caldwell}(2021)}]{Sabla:2021nfy}
\bibinfo{author}{\bibfnamefont{V.~I.} \bibnamefont{Sabla}} \bibnamefont{and}
  \bibinfo{author}{\bibfnamefont{R.~R.} \bibnamefont{Caldwell}}
  (\bibinfo{year}{2021}), \eprint{2103.04999}.

\bibitem[{\citenamefont{Nojiri et~al.}(2021)\citenamefont{Nojiri, Odintsov,
  Saez-Chillon~Gomez, and Sharov}}]{Nojiri:2021dze}
\bibinfo{author}{\bibfnamefont{S.}~\bibnamefont{Nojiri}},
  \bibinfo{author}{\bibfnamefont{S.~D.} \bibnamefont{Odintsov}},
  \bibinfo{author}{\bibfnamefont{D.}~\bibnamefont{Saez-Chillon~Gomez}},
  \bibnamefont{and} \bibinfo{author}{\bibfnamefont{G.~S.} \bibnamefont{Sharov}}
  (\bibinfo{year}{2021}), \eprint{2103.05304}.

\bibitem[{\citenamefont{Zumalacarregui}(2020)}]{Zumalacarregui:2020cjh}
\bibinfo{author}{\bibfnamefont{M.}~\bibnamefont{Zumalacarregui}},
  \bibinfo{journal}{Phys. Rev. D} \textbf{\bibinfo{volume}{102}},
  \bibinfo{pages}{023523} (\bibinfo{year}{2020}), \eprint{2003.06396}.

\bibitem[{\citenamefont{Ballesteros et~al.}(2020)\citenamefont{Ballesteros,
  Notari, and Rompineve}}]{Ballesteros:2020sik}
\bibinfo{author}{\bibfnamefont{G.}~\bibnamefont{Ballesteros}},
  \bibinfo{author}{\bibfnamefont{A.}~\bibnamefont{Notari}}, \bibnamefont{and}
  \bibinfo{author}{\bibfnamefont{F.}~\bibnamefont{Rompineve}},
  \bibinfo{journal}{JCAP} \textbf{\bibinfo{volume}{11}}, \bibinfo{pages}{024}
  (\bibinfo{year}{2020}), \eprint{2004.05049}.

\bibitem[{\citenamefont{Braglia
  et~al.}(2020{\natexlab{b}})\citenamefont{Braglia, Emond, Finelli,
  Gumrukcuoglu, and Koyama}}]{Braglia:2020bym}
\bibinfo{author}{\bibfnamefont{M.}~\bibnamefont{Braglia}},
  \bibinfo{author}{\bibfnamefont{W.~T.} \bibnamefont{Emond}},
  \bibinfo{author}{\bibfnamefont{F.}~\bibnamefont{Finelli}},
  \bibinfo{author}{\bibfnamefont{A.~E.} \bibnamefont{Gumrukcuoglu}},
  \bibnamefont{and} \bibinfo{author}{\bibfnamefont{K.}~\bibnamefont{Koyama}},
  \bibinfo{journal}{Phys. Rev. D} \textbf{\bibinfo{volume}{102}},
  \bibinfo{pages}{083513} (\bibinfo{year}{2020}{\natexlab{b}}),
  \eprint{2005.14053}.

\bibitem[{\citenamefont{Braglia
  et~al.}(2020{\natexlab{c}})\citenamefont{Braglia, Ballardini, Finelli, and
  Koyama}}]{Braglia:2020auw}
\bibinfo{author}{\bibfnamefont{M.}~\bibnamefont{Braglia}},
  \bibinfo{author}{\bibfnamefont{M.}~\bibnamefont{Ballardini}},
  \bibinfo{author}{\bibfnamefont{F.}~\bibnamefont{Finelli}}, \bibnamefont{and}
  \bibinfo{author}{\bibfnamefont{K.}~\bibnamefont{Koyama}}
  (\bibinfo{year}{2020}{\natexlab{c}}), \eprint{2011.12934}.

\bibitem[{\citenamefont{Odintsov et~al.}(2020)\citenamefont{Odintsov, G\'omez,
  and Sharov}}]{Odintsov:2020qzd}
\bibinfo{author}{\bibfnamefont{S.~D.} \bibnamefont{Odintsov}},
  \bibinfo{author}{\bibfnamefont{D.~S.-C.} \bibnamefont{G\'omez}},
  \bibnamefont{and} \bibinfo{author}{\bibfnamefont{G.~S.} \bibnamefont{Sharov}}
  (\bibinfo{year}{2020}), \eprint{2011.03957}.

\bibitem[{\citenamefont{Lemos et~al.}(2019)\citenamefont{Lemos, Lee,
  Efstathiou, and Gratton}}]{Lemos:2018smw}
\bibinfo{author}{\bibfnamefont{P.}~\bibnamefont{Lemos}},
  \bibinfo{author}{\bibfnamefont{E.}~\bibnamefont{Lee}},
  \bibinfo{author}{\bibfnamefont{G.}~\bibnamefont{Efstathiou}},
  \bibnamefont{and} \bibinfo{author}{\bibfnamefont{S.}~\bibnamefont{Gratton}},
  \bibinfo{journal}{Mon. Not. Roy. Astron. Soc.}
  \textbf{\bibinfo{volume}{483}}, \bibinfo{pages}{4803} (\bibinfo{year}{2019}),
  \eprint{1806.06781}.

\bibitem[{\citenamefont{Efstathiou}(2021)}]{Efstathiou:2021ocp}
\bibinfo{author}{\bibfnamefont{G.}~\bibnamefont{Efstathiou}}
  (\bibinfo{year}{2021}), \eprint{2103.08723}.

\bibitem[{\citenamefont{Vagnozzi}(2020)}]{Vagnozzi:2019ezj}
\bibinfo{author}{\bibfnamefont{S.}~\bibnamefont{Vagnozzi}},
  \bibinfo{journal}{Phys. Rev. D} \textbf{\bibinfo{volume}{102}},
  \bibinfo{pages}{023518} (\bibinfo{year}{2020}), \eprint{1907.07569}.

\bibitem[{\citenamefont{Di~Valentino
  et~al.}(2020{\natexlab{b}})\citenamefont{Di~Valentino, Melchiorri, Mena, and
  Vagnozzi}}]{DiValentino:2019jae}
\bibinfo{author}{\bibfnamefont{E.}~\bibnamefont{Di~Valentino}},
  \bibinfo{author}{\bibfnamefont{A.}~\bibnamefont{Melchiorri}},
  \bibinfo{author}{\bibfnamefont{O.}~\bibnamefont{Mena}}, \bibnamefont{and}
  \bibinfo{author}{\bibfnamefont{S.}~\bibnamefont{Vagnozzi}},
  \bibinfo{journal}{Phys. Rev. D} \textbf{\bibinfo{volume}{101}},
  \bibinfo{pages}{063502} (\bibinfo{year}{2020}{\natexlab{b}}),
  \eprint{1910.09853}.

\bibitem[{\citenamefont{Yang et~al.}(2020)\citenamefont{Yang, Di~Valentino,
  Pan, Basilakos, and Paliathanasis}}]{Yang:2020zuk}
\bibinfo{author}{\bibfnamefont{W.}~\bibnamefont{Yang}},
  \bibinfo{author}{\bibfnamefont{E.}~\bibnamefont{Di~Valentino}},
  \bibinfo{author}{\bibfnamefont{S.}~\bibnamefont{Pan}},
  \bibinfo{author}{\bibfnamefont{S.}~\bibnamefont{Basilakos}},
  \bibnamefont{and}
  \bibinfo{author}{\bibfnamefont{A.}~\bibnamefont{Paliathanasis}},
  \bibinfo{journal}{Phys. Rev. D} \textbf{\bibinfo{volume}{102}},
  \bibinfo{pages}{063503} (\bibinfo{year}{2020}), \eprint{2001.04307}.

\bibitem[{\citenamefont{Yang et~al.}(2021{\natexlab{a}})\citenamefont{Yang,
  Pan, Valentino, Mena, and Melchiorri}}]{yang20212021h0}
\bibinfo{author}{\bibfnamefont{W.}~\bibnamefont{Yang}},
  \bibinfo{author}{\bibfnamefont{S.}~\bibnamefont{Pan}},
  \bibinfo{author}{\bibfnamefont{E.~D.} \bibnamefont{Valentino}},
  \bibinfo{author}{\bibfnamefont{O.}~\bibnamefont{Mena}}, \bibnamefont{and}
  \bibinfo{author}{\bibfnamefont{A.}~\bibnamefont{Melchiorri}}
  (\bibinfo{year}{2021}{\natexlab{a}}), \eprint{2101.03129}.

\bibitem[{\citenamefont{Yang et~al.}(2021{\natexlab{b}})\citenamefont{Yang,
  Di~Valentino, Pan, Shafieloo, and Li}}]{Yang:2021eud}
\bibinfo{author}{\bibfnamefont{W.}~\bibnamefont{Yang}},
  \bibinfo{author}{\bibfnamefont{E.}~\bibnamefont{Di~Valentino}},
  \bibinfo{author}{\bibfnamefont{S.}~\bibnamefont{Pan}},
  \bibinfo{author}{\bibfnamefont{A.}~\bibnamefont{Shafieloo}},
  \bibnamefont{and} \bibinfo{author}{\bibfnamefont{X.}~\bibnamefont{Li}}
  (\bibinfo{year}{2021}{\natexlab{b}}), \eprint{2103.03815}.

\bibitem[{\citenamefont{Benetti et~al.}(2013)\citenamefont{Benetti, Gerbino,
  Kinney, Kolb, Lattanzi, Melchiorri, Pagano, and Riotto}}]{Benetti:2013wla}
\bibinfo{author}{\bibfnamefont{M.}~\bibnamefont{Benetti}},
  \bibinfo{author}{\bibfnamefont{M.}~\bibnamefont{Gerbino}},
  \bibinfo{author}{\bibfnamefont{W.~H.} \bibnamefont{Kinney}},
  \bibinfo{author}{\bibfnamefont{E.~W.} \bibnamefont{Kolb}},
  \bibinfo{author}{\bibfnamefont{M.}~\bibnamefont{Lattanzi}},
  \bibinfo{author}{\bibfnamefont{A.}~\bibnamefont{Melchiorri}},
  \bibinfo{author}{\bibfnamefont{L.}~\bibnamefont{Pagano}}, \bibnamefont{and}
  \bibinfo{author}{\bibfnamefont{A.}~\bibnamefont{Riotto}},
  \bibinfo{journal}{JCAP} \textbf{\bibinfo{volume}{10}}, \bibinfo{pages}{030}
  (\bibinfo{year}{2013}), \eprint{1303.4317}.

\bibitem[{\citenamefont{Gerbino et~al.}(2017)\citenamefont{Gerbino, Freese,
  Vagnozzi, Lattanzi, Mena, Giusarma, and Ho}}]{Gerbino:2016sgw}
\bibinfo{author}{\bibfnamefont{M.}~\bibnamefont{Gerbino}},
  \bibinfo{author}{\bibfnamefont{K.}~\bibnamefont{Freese}},
  \bibinfo{author}{\bibfnamefont{S.}~\bibnamefont{Vagnozzi}},
  \bibinfo{author}{\bibfnamefont{M.}~\bibnamefont{Lattanzi}},
  \bibinfo{author}{\bibfnamefont{O.}~\bibnamefont{Mena}},
  \bibinfo{author}{\bibfnamefont{E.}~\bibnamefont{Giusarma}}, \bibnamefont{and}
  \bibinfo{author}{\bibfnamefont{S.}~\bibnamefont{Ho}}, \bibinfo{journal}{Phys.
  Rev. D} \textbf{\bibinfo{volume}{95}}, \bibinfo{pages}{043512}
  (\bibinfo{year}{2017}), \eprint{1610.08830}.

\bibitem[{\citenamefont{Zhang}(2017)}]{Zhang:2017epd}
\bibinfo{author}{\bibfnamefont{X.}~\bibnamefont{Zhang}}, \bibinfo{journal}{Sci.
  China Phys. Mech. Astron.} \textbf{\bibinfo{volume}{60}},
  \bibinfo{pages}{060421} (\bibinfo{year}{2017}), \eprint{1702.05010}.

\bibitem[{\citenamefont{Benetti et~al.}(2017)\citenamefont{Benetti, Graef, and
  Alcaniz}}]{Benetti:2017gvm}
\bibinfo{author}{\bibfnamefont{M.}~\bibnamefont{Benetti}},
  \bibinfo{author}{\bibfnamefont{L.~L.} \bibnamefont{Graef}}, \bibnamefont{and}
  \bibinfo{author}{\bibfnamefont{J.~S.} \bibnamefont{Alcaniz}},
  \bibinfo{journal}{JCAP} \textbf{\bibinfo{volume}{04}}, \bibinfo{pages}{003}
  (\bibinfo{year}{2017}), \eprint{1702.06509}.

\bibitem[{\citenamefont{Benetti et~al.}(2018)\citenamefont{Benetti, Graef, and
  Alcaniz}}]{Benetti:2017juy}
\bibinfo{author}{\bibfnamefont{M.}~\bibnamefont{Benetti}},
  \bibinfo{author}{\bibfnamefont{L.~L.} \bibnamefont{Graef}}, \bibnamefont{and}
  \bibinfo{author}{\bibfnamefont{J.~S.} \bibnamefont{Alcaniz}},
  \bibinfo{journal}{JCAP} \textbf{\bibinfo{volume}{07}}, \bibinfo{pages}{066}
  (\bibinfo{year}{2018}), \eprint{1712.00677}.

\bibitem[{\citenamefont{Chiang and Slosar}(2018)}]{Chiang:2018xpn}
\bibinfo{author}{\bibfnamefont{C.-T.} \bibnamefont{Chiang}} \bibnamefont{and}
  \bibinfo{author}{\bibfnamefont{A.}~\bibnamefont{Slosar}}
  (\bibinfo{year}{2018}), \eprint{1811.03624}.

\bibitem[{\citenamefont{Hart and Chluba}(2020)}]{Hart:2019dxi}
\bibinfo{author}{\bibfnamefont{L.}~\bibnamefont{Hart}} \bibnamefont{and}
  \bibinfo{author}{\bibfnamefont{J.}~\bibnamefont{Chluba}},
  \bibinfo{journal}{Mon. Not. Roy. Astron. Soc.}
  \textbf{\bibinfo{volume}{493}}, \bibinfo{pages}{3255} (\bibinfo{year}{2020}),
  \eprint{1912.03986}.

\bibitem[{\citenamefont{Sekiguchi and Takahashi}(2020)}]{Sekiguchi:2020teg}
\bibinfo{author}{\bibfnamefont{T.}~\bibnamefont{Sekiguchi}} \bibnamefont{and}
  \bibinfo{author}{\bibfnamefont{T.}~\bibnamefont{Takahashi}}
  (\bibinfo{year}{2020}), \eprint{2007.03381}.

\bibitem[{\citenamefont{Jedamzik and Pogosian}(2020)}]{Jedamzik:2020krr}
\bibinfo{author}{\bibfnamefont{K.}~\bibnamefont{Jedamzik}} \bibnamefont{and}
  \bibinfo{author}{\bibfnamefont{L.}~\bibnamefont{Pogosian}},
  \bibinfo{journal}{Phys. Rev. Lett.} \textbf{\bibinfo{volume}{125}},
  \bibinfo{pages}{181302} (\bibinfo{year}{2020}), \eprint{2004.09487}.

\bibitem[{\citenamefont{Ade et~al.}(2016)}]{Array:2015xqh}
\bibinfo{author}{\bibfnamefont{P.~A.~R.} \bibnamefont{Ade}}
  \bibnamefont{et~al.} (\bibinfo{collaboration}{BICEP2, Keck Array}),
  \bibinfo{journal}{Phys. Rev. Lett.} \textbf{\bibinfo{volume}{116}},
  \bibinfo{pages}{031302} (\bibinfo{year}{2016}), \eprint{1510.09217}.

\bibitem[{\citenamefont{Harrison}(1970)}]{Harrison:1969fb}
\bibinfo{author}{\bibfnamefont{E.~R.} \bibnamefont{Harrison}},
  \bibinfo{journal}{Phys. Rev. D} \textbf{\bibinfo{volume}{1}},
  \bibinfo{pages}{2726} (\bibinfo{year}{1970}).

\bibitem[{\citenamefont{Zeldovich}(1972)}]{Zeldovich:1972zz}
\bibinfo{author}{\bibfnamefont{Y.}~\bibnamefont{Zeldovich}},
  \bibinfo{journal}{Mon. Not. Roy. Astron. Soc.}
  \textbf{\bibinfo{volume}{160}}, \bibinfo{pages}{1P} (\bibinfo{year}{1972}).

\bibitem[{\citenamefont{Peebles and Yu}(1970)}]{Peebles:1970ag}
\bibinfo{author}{\bibfnamefont{P.}~\bibnamefont{Peebles}} \bibnamefont{and}
  \bibinfo{author}{\bibfnamefont{J.}~\bibnamefont{Yu}},
  \bibinfo{journal}{Astrophys. J.} \textbf{\bibinfo{volume}{162}},
  \bibinfo{pages}{815} (\bibinfo{year}{1970}).

\bibitem[{\citenamefont{Akrami et~al.}(2020)}]{Akrami:2018odb}
\bibinfo{author}{\bibfnamefont{Y.}~\bibnamefont{Akrami}} \bibnamefont{et~al.}
  (\bibinfo{collaboration}{Planck}), \bibinfo{journal}{Astron. Astrophys.}
  \textbf{\bibinfo{volume}{641}}, \bibinfo{pages}{A10} (\bibinfo{year}{2020}),
  \eprint{1807.06211}.

\bibitem[{\citenamefont{Di~Valentino et~al.}(2018)\citenamefont{Di~Valentino,
  Melchiorri, Fantaye, and Heavens}}]{DiValentino:2018zjj}
\bibinfo{author}{\bibfnamefont{E.}~\bibnamefont{Di~Valentino}},
  \bibinfo{author}{\bibfnamefont{A.}~\bibnamefont{Melchiorri}},
  \bibinfo{author}{\bibfnamefont{Y.}~\bibnamefont{Fantaye}}, \bibnamefont{and}
  \bibinfo{author}{\bibfnamefont{A.}~\bibnamefont{Heavens}},
  \bibinfo{journal}{Phys. Rev. D} \textbf{\bibinfo{volume}{98}},
  \bibinfo{pages}{063508} (\bibinfo{year}{2018}), \eprint{1808.09201}.

\bibitem[{\citenamefont{Pogosian et~al.}(2020)\citenamefont{Pogosian, Zhao, and
  Jedamzik}}]{Pogosian:2020ded}
\bibinfo{author}{\bibfnamefont{L.}~\bibnamefont{Pogosian}},
  \bibinfo{author}{\bibfnamefont{G.-B.} \bibnamefont{Zhao}}, \bibnamefont{and}
  \bibinfo{author}{\bibfnamefont{K.}~\bibnamefont{Jedamzik}},
  \bibinfo{journal}{Astrophys. J. Lett.} \textbf{\bibinfo{volume}{904}},
  \bibinfo{pages}{L17} (\bibinfo{year}{2020}), \eprint{2009.08455}.

\bibitem[{\citenamefont{Hu et~al.}(1997)\citenamefont{Hu, Sugiyama, and
  Silk}}]{Hu:1995kot}
\bibinfo{author}{\bibfnamefont{W.}~\bibnamefont{Hu}},
  \bibinfo{author}{\bibfnamefont{N.}~\bibnamefont{Sugiyama}}, \bibnamefont{and}
  \bibinfo{author}{\bibfnamefont{J.}~\bibnamefont{Silk}},
  \bibinfo{journal}{Nature} \textbf{\bibinfo{volume}{386}}, \bibinfo{pages}{37}
  (\bibinfo{year}{1997}), \eprint{astro-ph/9604166}.

\bibitem[{\citenamefont{Fixsen et~al.}(1996)\citenamefont{Fixsen, Cheng, Gales,
  Mather, Shafer, and Wright}}]{Fixsen:1996nj}
\bibinfo{author}{\bibfnamefont{D.~J.} \bibnamefont{Fixsen}},
  \bibinfo{author}{\bibfnamefont{E.~S.} \bibnamefont{Cheng}},
  \bibinfo{author}{\bibfnamefont{J.~M.} \bibnamefont{Gales}},
  \bibinfo{author}{\bibfnamefont{J.~C.} \bibnamefont{Mather}},
  \bibinfo{author}{\bibfnamefont{R.~A.} \bibnamefont{Shafer}},
  \bibnamefont{and} \bibinfo{author}{\bibfnamefont{E.~L.}
  \bibnamefont{Wright}}, \bibinfo{journal}{Astrophys. J.}
  \textbf{\bibinfo{volume}{473}}, \bibinfo{pages}{576} (\bibinfo{year}{1996}),
  \eprint{astro-ph/9605054}.

\bibitem[{\citenamefont{Fixsen}(2009)}]{Fixsen:2009ug}
\bibinfo{author}{\bibfnamefont{D.~J.} \bibnamefont{Fixsen}},
  \bibinfo{journal}{Astrophys. J.} \textbf{\bibinfo{volume}{707}},
  \bibinfo{pages}{916} (\bibinfo{year}{2009}), \eprint{0911.1955}.

\bibitem[{\citenamefont{Riess et~al.}(2019)\citenamefont{Riess, Casertano,
  Yuan, Macri, and Scolnic}}]{Riess:2019cxk}
\bibinfo{author}{\bibfnamefont{A.~G.} \bibnamefont{Riess}},
  \bibinfo{author}{\bibfnamefont{S.}~\bibnamefont{Casertano}},
  \bibinfo{author}{\bibfnamefont{W.}~\bibnamefont{Yuan}},
  \bibinfo{author}{\bibfnamefont{L.~M.} \bibnamefont{Macri}}, \bibnamefont{and}
  \bibinfo{author}{\bibfnamefont{D.}~\bibnamefont{Scolnic}},
  \bibinfo{journal}{Astrophys. J.} \textbf{\bibinfo{volume}{876}},
  \bibinfo{pages}{85} (\bibinfo{year}{2019}), \eprint{1903.07603}.

\bibitem[{\citenamefont{Alam et~al.}(2017)}]{Alam:2016hwk}
\bibinfo{author}{\bibfnamefont{S.}~\bibnamefont{Alam}} \bibnamefont{et~al.}
  (\bibinfo{collaboration}{BOSS}), \bibinfo{journal}{Mon. Not. Roy. Astron.
  Soc.} \textbf{\bibinfo{volume}{470}}, \bibinfo{pages}{2617}
  (\bibinfo{year}{2017}), \eprint{1607.03155}.

\bibitem[{\citenamefont{Beutler et~al.}(2011)\citenamefont{Beutler, Blake,
  Colless, Jones, Staveley-Smith, Campbell, Parker, Saunders, and
  Watson}}]{Beutler:2011hx}
\bibinfo{author}{\bibfnamefont{F.}~\bibnamefont{Beutler}},
  \bibinfo{author}{\bibfnamefont{C.}~\bibnamefont{Blake}},
  \bibinfo{author}{\bibfnamefont{M.}~\bibnamefont{Colless}},
  \bibinfo{author}{\bibfnamefont{D.~H.} \bibnamefont{Jones}},
  \bibinfo{author}{\bibfnamefont{L.}~\bibnamefont{Staveley-Smith}},
  \bibinfo{author}{\bibfnamefont{L.}~\bibnamefont{Campbell}},
  \bibinfo{author}{\bibfnamefont{Q.}~\bibnamefont{Parker}},
  \bibinfo{author}{\bibfnamefont{W.}~\bibnamefont{Saunders}}, \bibnamefont{and}
  \bibinfo{author}{\bibfnamefont{F.}~\bibnamefont{Watson}},
  \bibinfo{journal}{Mon. Not. Roy. Astron. Soc.}
  \textbf{\bibinfo{volume}{416}}, \bibinfo{pages}{3017} (\bibinfo{year}{2011}),
  \eprint{1106.3366}.

\bibitem[{\citenamefont{Ross et~al.}(2015)\citenamefont{Ross, Samushia,
  Howlett, Percival, Burden, and Manera}}]{Ross:2014qpa}
\bibinfo{author}{\bibfnamefont{A.~J.} \bibnamefont{Ross}},
  \bibinfo{author}{\bibfnamefont{L.}~\bibnamefont{Samushia}},
  \bibinfo{author}{\bibfnamefont{C.}~\bibnamefont{Howlett}},
  \bibinfo{author}{\bibfnamefont{W.~J.} \bibnamefont{Percival}},
  \bibinfo{author}{\bibfnamefont{A.}~\bibnamefont{Burden}}, \bibnamefont{and}
  \bibinfo{author}{\bibfnamefont{M.}~\bibnamefont{Manera}},
  \bibinfo{journal}{Mon. Not. Roy. Astron. Soc.}
  \textbf{\bibinfo{volume}{449}}, \bibinfo{pages}{835} (\bibinfo{year}{2015}),
  \eprint{1409.3242}.

\bibitem[{\citenamefont{Scolnic et~al.}(2018)}]{Scolnic:2017caz}
\bibinfo{author}{\bibfnamefont{D.~M.} \bibnamefont{Scolnic}}
  \bibnamefont{et~al.}, \bibinfo{journal}{Astrophys. J.}
  \textbf{\bibinfo{volume}{859}}, \bibinfo{pages}{101} (\bibinfo{year}{2018}),
  \eprint{1710.00845}.

\bibitem[{\citenamefont{Audren et~al.}(2013)\citenamefont{Audren, Lesgourgues,
  Benabed, and Prunet}}]{Audren:2012wb}
\bibinfo{author}{\bibfnamefont{B.}~\bibnamefont{Audren}},
  \bibinfo{author}{\bibfnamefont{J.}~\bibnamefont{Lesgourgues}},
  \bibinfo{author}{\bibfnamefont{K.}~\bibnamefont{Benabed}}, \bibnamefont{and}
  \bibinfo{author}{\bibfnamefont{S.}~\bibnamefont{Prunet}},
  \bibinfo{journal}{JCAP} \textbf{\bibinfo{volume}{02}}, \bibinfo{pages}{001}
  (\bibinfo{year}{2013}), \eprint{1210.7183}.

\bibitem[{\citenamefont{Brinckmann and Lesgourgues}(2019)}]{Brinckmann:2018cvx}
\bibinfo{author}{\bibfnamefont{T.}~\bibnamefont{Brinckmann}} \bibnamefont{and}
  \bibinfo{author}{\bibfnamefont{J.}~\bibnamefont{Lesgourgues}},
  \bibinfo{journal}{Phys. Dark Univ.} \textbf{\bibinfo{volume}{24}},
  \bibinfo{pages}{100260} (\bibinfo{year}{2019}), \eprint{1804.07261}.

\bibitem[{\citenamefont{Lesgourgues}(2011)}]{Lesgourgues:2011re}
\bibinfo{author}{\bibfnamefont{J.}~\bibnamefont{Lesgourgues}}
  (\bibinfo{year}{2011}), \eprint{1104.2932}.

\bibitem[{\citenamefont{Blas et~al.}(2011)\citenamefont{Blas, Lesgourgues, and
  Tram}}]{Blas:2011rf}
\bibinfo{author}{\bibfnamefont{D.}~\bibnamefont{Blas}},
  \bibinfo{author}{\bibfnamefont{J.}~\bibnamefont{Lesgourgues}},
  \bibnamefont{and} \bibinfo{author}{\bibfnamefont{T.}~\bibnamefont{Tram}},
  \bibinfo{journal}{JCAP} \textbf{\bibinfo{volume}{07}}, \bibinfo{pages}{034}
  (\bibinfo{year}{2011}), \eprint{1104.2933}.

\bibitem[{\citenamefont{Lucca}(2020)}]{Lucca:2020fgp}
\bibinfo{author}{\bibfnamefont{M.}~\bibnamefont{Lucca}},
  \bibinfo{journal}{Phys. Lett. B} \textbf{\bibinfo{volume}{810}},
  \bibinfo{pages}{135791} (\bibinfo{year}{2020}), \eprint{2008.01115}.

\bibitem[{\citenamefont{Krishnan et~al.}(2020)\citenamefont{Krishnan, Colgain,
  Sheikh-Jabbari, and Yang}}]{Krishnan:2020vaf}
\bibinfo{author}{\bibfnamefont{C.}~\bibnamefont{Krishnan}},
  \bibinfo{author}{\bibfnamefont{E.~O.} \bibnamefont{Colgain}},
  \bibinfo{author}{\bibfnamefont{M.~M.} \bibnamefont{Sheikh-Jabbari}},
  \bibnamefont{and} \bibinfo{author}{\bibfnamefont{T.}~\bibnamefont{Yang}}
  (\bibinfo{year}{2020}), \eprint{2011.02858}.

\bibitem[{\citenamefont{Bernal et~al.}(2021)\citenamefont{Bernal, Verde,
  Jimenez, Kamionkowski, Valcin, and Wandelt}}]{Bernal:2021yli}
\bibinfo{author}{\bibfnamefont{J.~L.} \bibnamefont{Bernal}},
  \bibinfo{author}{\bibfnamefont{L.}~\bibnamefont{Verde}},
  \bibinfo{author}{\bibfnamefont{R.}~\bibnamefont{Jimenez}},
  \bibinfo{author}{\bibfnamefont{M.}~\bibnamefont{Kamionkowski}},
  \bibinfo{author}{\bibfnamefont{D.}~\bibnamefont{Valcin}}, \bibnamefont{and}
  \bibinfo{author}{\bibfnamefont{B.~D.} \bibnamefont{Wandelt}}
  (\bibinfo{year}{2021}), \eprint{2102.05066}.

\bibitem[{\citenamefont{Fanizza et~al.}(2021)\citenamefont{Fanizza, Fiorini,
  and Marozzi}}]{Fanizza:2021tuh}
\bibinfo{author}{\bibfnamefont{G.}~\bibnamefont{Fanizza}},
  \bibinfo{author}{\bibfnamefont{B.}~\bibnamefont{Fiorini}}, \bibnamefont{and}
  \bibinfo{author}{\bibfnamefont{G.}~\bibnamefont{Marozzi}}
  (\bibinfo{year}{2021}), \eprint{2102.12419}.

\bibitem[{\citenamefont{Henning et~al.}(2018)}]{Henning:2017nuy}
\bibinfo{author}{\bibfnamefont{J.~W.} \bibnamefont{Henning}}
  \bibnamefont{et~al.} (\bibinfo{collaboration}{SPT}),
  \bibinfo{journal}{Astrophys. J.} \textbf{\bibinfo{volume}{852}},
  \bibinfo{pages}{97} (\bibinfo{year}{2018}), \eprint{1707.09353}.

\bibitem[{\citenamefont{Choi et~al.}(2020)}]{Choi:2020ccd}
\bibinfo{author}{\bibfnamefont{S.~K.} \bibnamefont{Choi}} \bibnamefont{et~al.}
  (\bibinfo{collaboration}{ACT}), \bibinfo{journal}{JCAP}
  \textbf{\bibinfo{volume}{12}}, \bibinfo{pages}{045} (\bibinfo{year}{2020}),
  \eprint{2007.07289}.

\end{thebibliography}
\end{document}